\documentclass[letterpaper,twocolumn,prl, showpacs]{revtex4}
\usepackage{graphicx, amsmath,amssymb, multirow}

\newcommand{\be}[0]{\begin{equation}}
\newcommand{\ee}[0]{\end{equation}}
\newcommand{\bea}[0]{\begin{eqnarray}}
\newcommand{\eea}[0]{\end{eqnarray}}

\begin{document}

\title{Measurement of the Orbital Angular Momentum Spectrum of Fields with Partial Angular Coherence using Double Angular Slit Interference}
\author{Mehul Malik$^1$}
\author{Sangeeta Murugkar$^2$}
\author{Jonathan Leach$^2$}
\author{Robert W. Boyd$^{1,2}$}
\affiliation{$^1$The Institute of Optics, University of Rochester,
Rochester, New York 14627, USA}
\affiliation{$^2$Department of Physics, University of Ottawa, 150 Louis Pasteur,
Ottawa, ON, K1N~6N5, Canada}
\email{memalik@optics.rochester.edu}
\date{\today}

\begin{abstract}

We implement an interferometric method using two angular slits to measure the orbital angular momentum (OAM) mode spectrum of a field with partial angular coherence. As the angular separation of the slits changes, an interference pattern for a particular OAM mode is obtained. The visibility of this interference pattern as a function of angular separation is equivalent to the angular correlation function of the field. By Fourier transforming the angular correlation function obtained from the double angular slit interference, we are able to calculate the OAM spectrum of the partially coherent field. This method has potential application for characterizing the OAM spectrum in high-dimensional quantum information protocols.



\end{abstract}


\maketitle

\section{I. Introduction}\label{Intro}

Historically, the problem of determining the correlations between two space-time points in a wave field has been addressed by the analysis of a two-beam interference experiment \cite{Mandel:1995,Dixon:2010kz}. The degree of spatial coherence of a light source can be determined from the visibility of fringes in the far-field, arising from the interference of light fields in a Young's double slit experiment. The necessary condition for interference is that the slit separation should be smaller than the transverse coherence length of the field.  

It is well known that angular position and its conjugate variable, orbital angular momentum (OAM), form Fourier transform pairs \cite{Yao:2006vc}. This is convincingly demonstrated by the experimental observation of a discrete diffraction pattern in the OAM spectrum when light is transmitted through angular amplitude masks \cite{Jack:2008gx,Jha:2008vp}. Extending this idea, one can construct an angular analog of Young's double slit experiment with two angular slits \cite{Jha:2011dv}. By measuring the visibility of the interference pattern obtained from double angular slit interference, one can calculate the degree of angular coherence of the source. For certain types of partially coherent sources, the OAM mode spectrum is simply a Fourier transform of the degree of angular coherence \cite{Jha:2011dv}. This provides us with a new way of measuring the OAM mode spectrum of partially coherent fields.

Interest in the measurement and characterization of the OAM mode spectrum has grown rapidly over the last decade. This is because OAM modes, by virtue of living in a discrete, infinite-dimensional Hilbert space, are a very useful tool for quantum information science. Entanglement in up to 12 dimensions has been shown using OAM modes \cite{Dada:2011dn}. Efforts are under way to realize a quantum key distribution system using the OAM basis and the conjugate basis of angular position \cite{Boyd:2011ff,Malik:2012ka}. Recently, a terabit/sec free space communication system using OAM modes was demonstrated \cite{Wang:2012ha}. The effects of turbulence on quantum communication using OAM modes is currently an active topic of research \cite{Pors:2011iw}. OAM modes are also being used as a tool for alignment free quantum communication due to their rotational invariance \cite{DAmbrosio:2012tr}. It is clear that accurate methods of characterizing the OAM mode spectrum will be of great importance for such quantum communication systems employing OAM modes.

\begin{figure}[b!]
  \centering
 \includegraphics[]{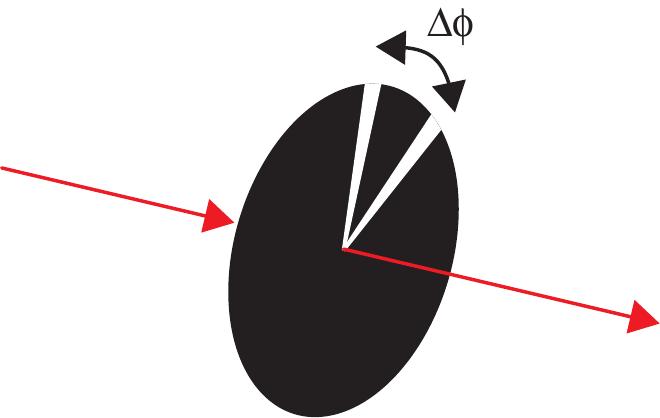}
 \caption{(Color online) Partially coherent field propagating through double angular slits with an angular separation of $\Delta\phi$.}\label{2slits}
\end{figure}

\begin{figure*}
  \centering
 \includegraphics[scale = 0.85]{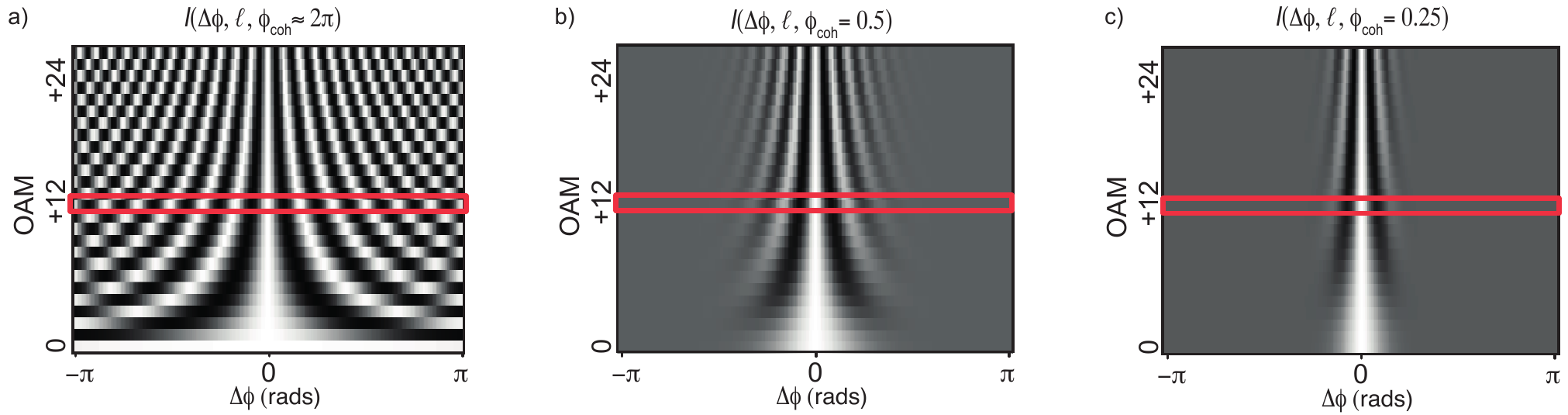}
 \caption{(Color online) Simulation results for angular interference patterns for OAM mode projections from $\ell=0$ to $\ell=24$ for three values of coherence angle (a) $\phi_{\textrm{coh}}\approx 2\pi$ rads (b) $\phi_{\textrm{coh}}$ = 0.5 rad and (c) $\phi_{\textrm{coh}}$ = 0.25 rad. The red rectangular boxes indicate the value of $\ell$ used in our experiment. The z-axis shows normalized intensity from 0 to 1, indicated by grayscale values.}\label{simresults}
\end{figure*}

The simplest way of measuring the OAM mode spectrum is by projection measurements of the constituent modes using a forked hologram \cite{Mair:2001ub}. More novel methods have been developed recently that involve interfering the field with a rotated copy of itself in a Mach-Zehnder interferometer \cite{Zambrini:2006jd,Pires:2010uz}. In this paper, we experimentally demonstrate a new way of measuring the OAM mode spectrum of fields with partial angular coherence that does not require careful interferometric alignment or image rotation. Based on the theoretical work in Ref. \cite{Jha:2011dv}, we first review the concepts of the degree of angular coherence and double angular slit interference. Then, we describe our experimental implementation of double angular slit interference using a spatial light modulator. Finally, we calculate the OAM mode spectrum from the visibility of the measured angular interference pattern.  

\section{II. Theory}\label{Theory}

The cross-spectral-density function is used to quantify field correlations between two points in the space-frequency domain. The non-negative definiteness of this function, in conjunction with the Mercer theorem, imply that it is a Hilbert-Schmidt kernel and that it has a coherent-mode representation of the form \cite{Mandel:1995}

\be W(\textbf{r}_1,\textbf{r}_2) = \displaystyle\sum\limits_{n} \alpha_n \psi^*_{n}(\textbf{r}_1) \psi_n(\textbf{r}_2) \ee

\noindent The properties of $W(\textbf{r}_1,\textbf{r}_2)$ ensure that all the coefficients $\alpha_n$ are real and non-negative, i.e. $\alpha_n\geq 0$, and that at least one $\alpha_n$ is non-zero. This further implies that for any partially coherent field, there exists at least one basis in which the cross-spectral-density function can be represented as the superposition of modes that are completely coherent in the space-frequency domain.

For fields in the Laguerre-Gauss (LG) basis, one can define a cross-spectral-density function in the angle-frequency domain by integrating over the radial dimension \cite{Jha:2011dv}. The cross-spectral-density function then represents the correlation between the fields at angular positions $\phi_1$ and $\phi_2$ and can be expressed as a superposition of OAM modes

\be W(\phi_1,\phi_2)=\displaystyle\sum\limits_{{\ell}=-\infty}^\infty C_{\ell}e^{-i\ell\Delta\phi}\ee

\noindent where $\Delta\phi=\phi_1-\phi_2$. If there is only one term in the above expansion, the field can be considered to be completely coherent in the angle-frequency domain. If there is more than one term in the expansion, the field can be characterized as being partially coherent. For such fields, one can introduce the angular analog to coherence length, the coherence angle, $\phi_{\textrm{coh}}$. The coherence angle quantifies the angular separation over which two field points are mutually coherent. 

These concepts are best illustrated by the case of angular interference from a double angular slit (Fig. \ref{2slits}). When a partially coherent field is transmitted through a mask containing two angular slits at angular positions $\phi_1$ and $\phi_2$ with unit transmission, an interference pattern is obtained with intensity in an OAM mode $\ell$ given by \cite{Jha:2011dv}

\begin{equation}\label{equation}
I_{\ell} (\Delta \phi)=\frac{1}{\pi}
\displaystyle\sum\limits_{{\ell^\prime}=-\infty}^\infty C_{\ell^\prime}[1+ \lambda(\Delta\phi) cos(\ell\Delta\phi)] .
\end{equation}

\noindent Here, $\lambda(\Delta\phi)$ is the degree of angular coherence and is obtained by normalizing the cross spectral density function:

\begin{eqnarray}\label{equation}
\lambda(\Delta\phi)&=&\frac{W(\phi_1,\phi_2)}{[W(\phi_1,\phi_1)]^{1/2}[W(\phi_2,\phi_2)]^{1/2}}\nonumber\\
&=&\frac{W(\phi_1,\phi_2)}{\displaystyle\sum\limits_{{\ell^\prime}=-\infty}^\infty C_{\ell^\prime}} 
\end{eqnarray}

\noindent where $\Delta\phi = \phi_1-\phi_2$. Eq. 3 is the angular interference law, since it quantifies the interference between the fields coming from two separate angular positions. This is analogous to the general interference law for stationary optical fields \cite{Mandel:1995}. Eq. 4 is analogous to the law relating the degree of spatial coherence to the cross-spectral density function in the space-frequency domain (Eq. 1). Also, one can see that $\lambda(\Delta\phi)$ is equal to the visibility of the interference pattern, which is a function of the angular slit separation. The width of $\lambda(\Delta\phi)$ is a measure of the coherence angle ($\phi_{\textrm{coh}}$) over which the field remains coherent. For a completely coherent field, the degree of angular coherence $\lambda(\Delta\phi)$ is equal to unity for all values of $\Delta\phi$, and $\phi_{\textrm{coh}}$ approaches a value of $2\pi$. 

It has been shown in Ref. \cite{Jha:2011dv} that for partially coherent fields with a broad Gaussian distribution in $\ell$, the angular correlation function is simply the Fourier transform of the OAM mode distribution. More specifically, for a Gaussian OAM mode distribution with a width $\sigma$,

\begin{equation}\label{equation}
C(\ell)=\frac{1}{\sqrt{2\pi}\sigma}\exp \left( \frac{-\ell^2}{2\sigma^2} \right).
\end{equation}

For such a distribution of OAM, the degree of angular coherence is given by

\begin{equation}\label{equation}
\lambda(\Delta\phi)=\exp \left(\frac{-\sigma^2\Delta\phi^2}{2} \right).
\end{equation}

By measuring the visibility of the interference pattern as a function of slit separation $\Delta\phi$, we can construct the degree of coherence $\lambda(\Delta\phi)$ of the field. The width of the visibility function, $1/\sigma$, gives us the coherence angle of the field, $\phi_{\textrm{coh}}$. From this, we can calculate $\sigma$, which is the width of the OAM mode spectrum of the partially coherent field, and reconstruct the OAM mode distribution from Eq. (5).  

Simulation results for angular interference patterns for OAM mode projections from $\ell=0$ to $\ell=24$ are plotted in Fig. \ref{simresults} (a)-(c) for three different values of $\phi_{\textrm{coh}}$. It is clear from Eq. 6 that the degree of coherence $\lambda(\Delta\phi)$ is independent of the OAM mode number used for projection. Consequently, the width of the visibility envelope, i.e. the coherence angle ($\phi_{\textrm{coh}}$), is also independent of the OAM mode number used for projection. This is clearly illustrated in the simulated results in Fig. \ref{simresults}. The inverse relationship between the coherence angle ($\phi_{\textrm{coh}}$) and the OAM mode distribution width ($\sigma$) is also apparent. The simulated interference pattern obtained by projecting in OAM mode $\ell=12$ is outlined in red. In the next section, we discuss experimental results for this particular case.


\begin{figure}[t!]
  \centering
 \includegraphics[]{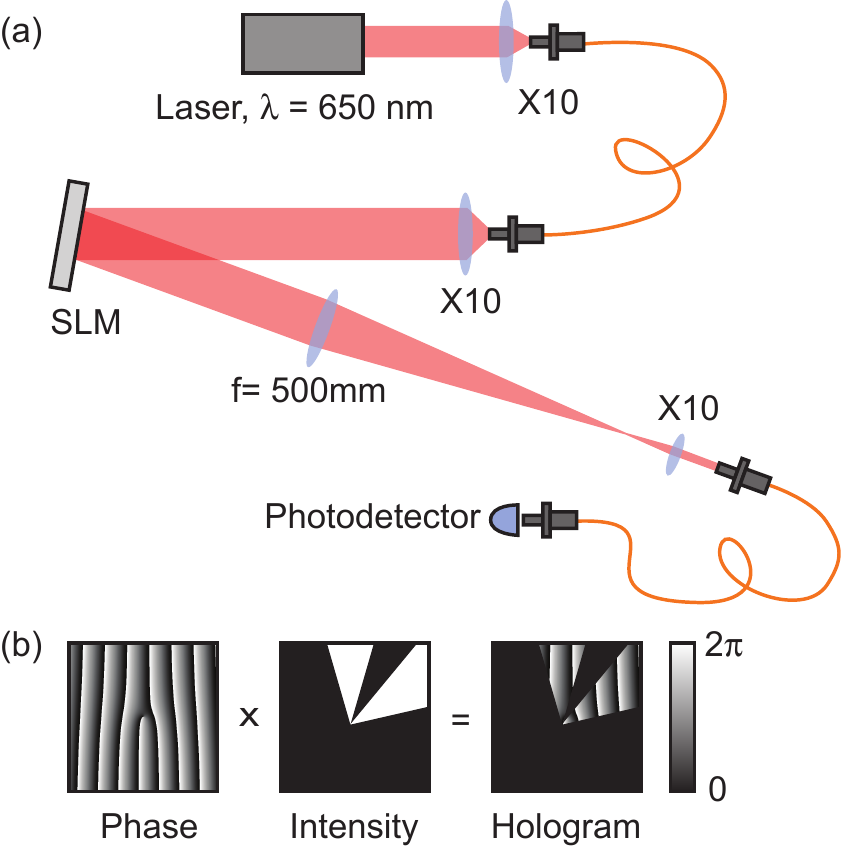}
 \caption{(Color online) (a) Schematic of the experimental setup (b) Example of a phase hologram impressed on the SLM}\label{setup}
\end{figure}

\begin{figure*}
  \centering
 \includegraphics[scale = 0.85]{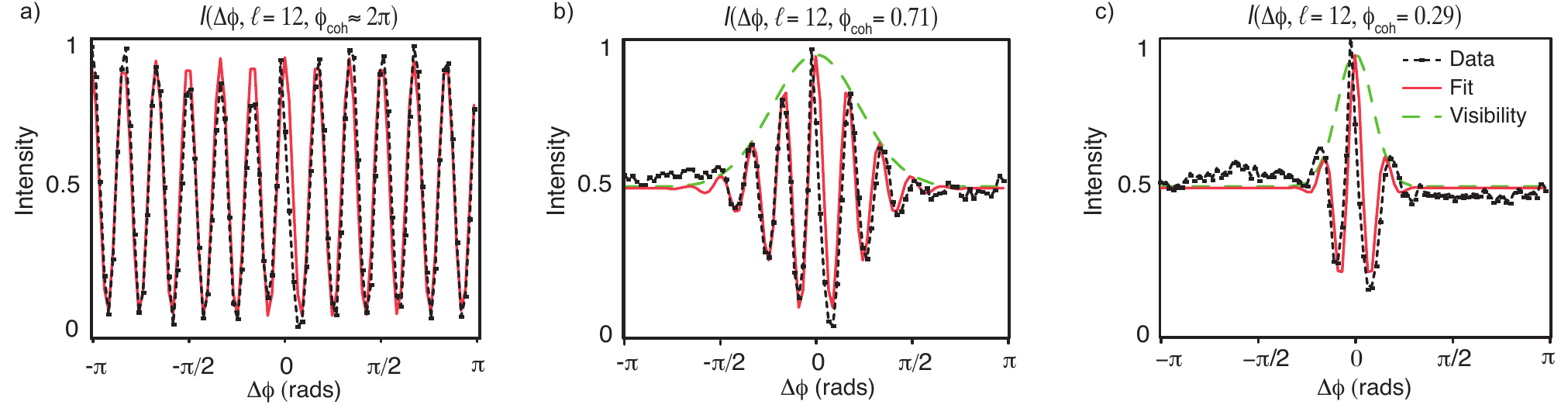}
 \caption{(Color online) Normalized intensity for OAM mode projection $\ell$=12, of partially coherent light transmitted through two angular apertures, as a function of the angular separation $\Delta\phi$ for three OAM mode distributions with different degrees of angular coherence, characterized by a coherence angle (a) $\phi_{\textrm{coh}}\approx 2\pi$ rads (b) $\phi_{\textrm{coh}}$ = 0.71 rad and (c) $\phi_{\textrm{coh}}$ = 0.29 rad. The measured interference patterns as a function of angular slit separation $\Delta\phi$ (dashed black line) are plotted alongside a fit (solid red line) obtained from theory. A fit to the visibility envelope is also plotted in (b) and (c) (long-dashed green line).}\label{finalresults}
\end{figure*}


\section{III. Experiment and Results}\label{Experiment and Results}

We implemented a double angular slit experiment using a setup shown in Fig. \ref{setup}(a). Light from a laser diode at 650nm is coupled into a single mode fiber (SMF) with a 10x microscope objective. The Gaussian single mode output from the SMF is collimated and is incident on the front surface of a spatial light modulator (SLM). It is a simple task to use an SLM to generate a coherent superposition
of OAM modes; however, for this experiment, we require partially coherent
Gaussian distributions of OAM modes with varying degrees of coherence.

To generate these distributions, we generate a coherent
superposition of OAM modes with a single hologram.  The range of
OAM modes that we generate at this stage has a width
$\sigma$, but each of the modes in the superposition is given a random
phase. We then generate a different superposition of OAM modes using a new
hologram in which each of the modes is given a new random phase. Changing the holograms in this way ensures that
there is no phase relationship between the modes generated by
subsequent holograms. This procedure is repeated over a thousand
times such that the resulting time-averaged beam that reflects off the
SLM is a partially coherent superposition of OAM modes with a coherence angle $\phi_{\textrm{coh}}\approx 1/\sigma$.

One should note that at this stage, we lack precise control over the degree of coherence of the partially coherent OAM mode distribution that is generated. We believe that a significant contributing factor to this effect is an $\ell$-dependent efficiency when using an SLM to generate modes that carry OAM. As SLMs are pixellated devices, and one requires an SLM to impart an azimuthal phase shift of $2 \pi \ell $ to generate a particular OAM mode, there is decreasing efficiency for generating modes with increasing OAM number. This can be understood as an $\ell$-dependent Nyquist limit when generating OAM modes with a pixellated device. For our experiment, the reduction in efficiency for large values of OAM results in generating a partially coherent mode distribution with a narrower width ($\sigma$) and larger coherence angle ($\phi_{\textrm{coh}}$) than intended. Calibration of this effect would improve the accuracy with which we can create mode distributions with a given degree of coherence. However, the strength of our proposed technique is that we have an accurate measure of the final OAM mode distribution width and coherence without relying on the $\ell$-dependent efficiency of an SLM.

For every partially coherent Gaussian distribution of OAM modes that we generate, the same SLM is then used to (i) generate a forked
phase hologram to project the field into an OAM mode with angular
momentum $\ell$, and (ii) generate an amplitude mask of two angular slits with $15^{\circ}$ widths, with one slit fixed at $0^{\circ}$ and the other rotating from $-\pi$ to $\pi$ radians. One should note that the same SLM is used to generate the partially coherent fields and carry out the angular interference experiment. An example of the phase and amplitude pattern impressed on the SLM for the angular interference experiment is shown in Fig. \ref{setup}(b). The light from the SLM is imaged onto the aperture of a SMF connected to a photodetector. 

The theory presented in the previous section assumes infinitesimal angular slits. Since this is not possible to implement in the lab, we use slits with a finite width of $15^{\circ}$. Analogous to the linear case, this results in a finite spread of OAM modes. In our experiment, we chose to measure the interference pattern given by Eq. (3) by projecting the field into a mode with $\ell$ = 12, which is present in the envelope of OAM modes diffracted from our finite width slits. The intensity at the photodetector as a function of the angular separation $\Delta\phi$ between the angular slits is shown in Fig. \ref{finalresults} (a)-(c). For a completely coherent field ($\phi_{\textrm{coh}}\approx 2\pi$), there is only a single OAM mode ($\ell=0$) in the generated field and the degree of coherence $\lambda(\Delta\phi)\approx 1$. The forked hologram with $\ell=12$ projects the field into an OAM mode with $\ell=-12$. Eq. (3) then simplifies to the form
\begin{equation}\label{equation}
I(\Delta \phi)=\frac{1}{\pi}
[1+ cos(12\Delta\phi)] .
\end{equation}

From this, we can see that the interference pattern thus obtained will have 12 fringes. This can be interpreted as there being 12 angular positions of the second angular slit as it rotates from $-\pi$ to $\pi$ radians, for which the fields transmitted through the two angular slits are exactly in phase. This gives rise to the 12 fringes seen in the intensity plot in Fig. \ref{finalresults}(a) (plotted with a dashed black line). A fit to these interference fringes based on the theory in section II is plotted with a solid red line. Since the field used in Fig. \ref{finalresults}(a) has only one OAM mode and is completely coherent in angular position, the visibility envelope of the fringe pattern is close to one for all values of angular separation. In Fig. \ref{finalresults}(b), the degree of coherence is reduced and there are additional OAM modes in the input Gaussian distribution. One can see that the visibility envelope is narrower, which leads to only 5 fringes being detected in the interference pattern. This is due to the fact that the angular width over which the field remains mutually coherent, or the coherence angle $\phi_{\textrm{coh}}$, is reduced. It can be seen from Fig. \ref{finalresults} (a)-(c) that as the width of the visibility envelope given by the coherence angle decreases further, fewer interference fringes are visible as $\Delta\phi$ goes from $-\pi$ to $\pi$ radians. 

\begin{figure}[t!]
  \centering
 \includegraphics[]{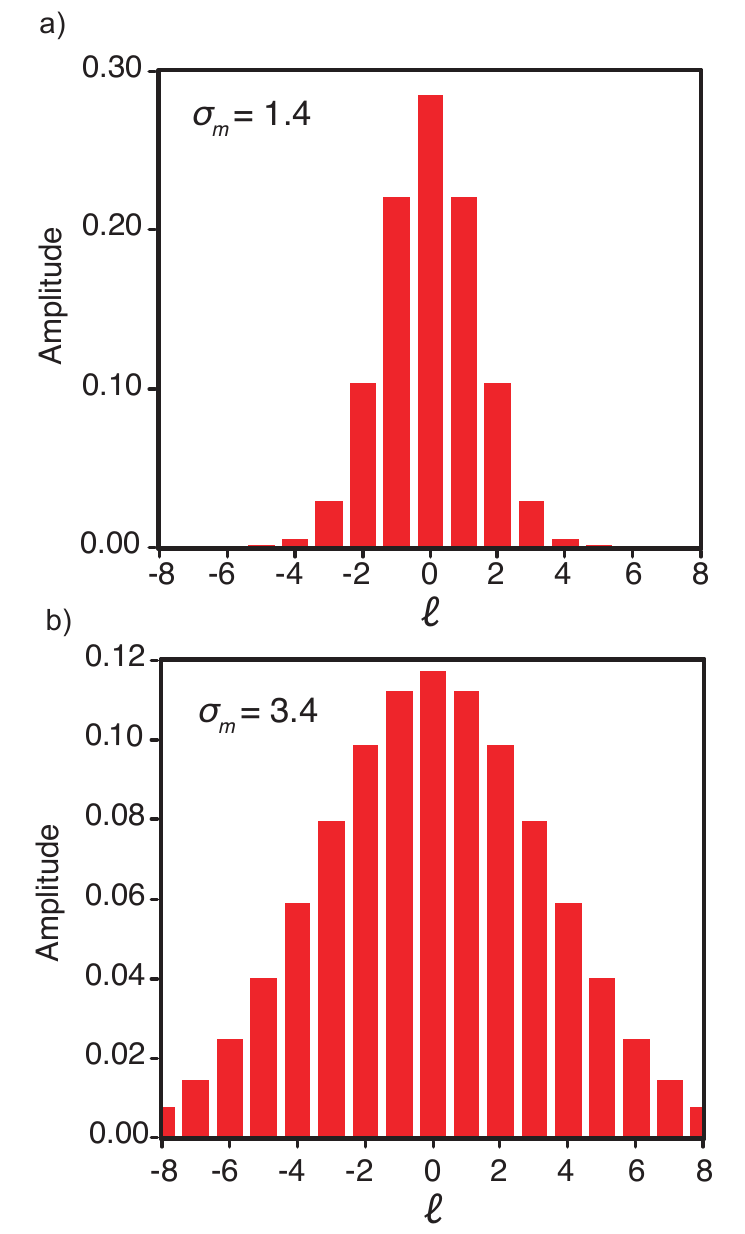}
 \caption{(Color online) Calculated OAM mode spectrum for the two measured distribution widths of (a) $\sigma_m= 1.4$ and (b) $\sigma_m = 3.4$. The normalized amplitude is plotted as a function of OAM mode number $\ell$.}\label{OAM_spectrum}
\end{figure}

When the input OAM mode distribution has a broad gaussian shape, we can plot the visibility envelope of the interference pattern using Eq. (6). This condition is held for the interference patterns seen in In Figs. \ref{finalresults} (b) and (c). We plot a fit to the visibility envelope (with a long-dashed green line) of the angular interference pattern for these two cases. The width of the visibility curve gives us the value of the coherence angle, $\phi_{\textrm{coh}}$. The measured value of $\phi_{\textrm{coh}}$ in Figs. \ref{finalresults} (b) and (c) is $0.71$ and $0.29$ radian respectively. From Eqs. (5) and (6), we see that the width of the OAM mode distribution is equal to the inverse of the coherence angle. Using this, we calculate the measured OAM mode distribution width $\sigma_m = 1.4$ and 3.4. The measured OAM mode distributions for these two cases are plotted in Fig. \ref{OAM_spectrum}. 

\section{IV. Conclusions}\label{Conclusions}

We have demonstrated experimental results to support the theoretical analysis of angular coherence proposed in Ref. \cite{Jha:2011dv} for a partially coherent light source. We employed an SLM and a classical light source to create a partially coherent Gaussian distribution of OAM modes. When this beam was passed through a double angular-slit, a characteristic interference pattern consisting of fringes inside a visibility envelope was obtained for a specific OAM mode projection. The width of this visibility envelope served as a direct measure of the degree of angular coherence of the source, and allowed us to reconstruct the input OAM mode distribution to a good approximation. This was consistent with our simulation results which were modeled on the theory of partial angular coherence put forth in Ref. \cite{Jha:2011dv}. We envision that such measurements will be useful for quantum information protocols that rely on the high-dimensionality of the OAM state space \cite{Groeblacher:2005ec,Boyd:2011ff}. As these protocols are usually implemented in free-space communication systems \cite{Malik:2012ka}, it is important to characterize the effects of turbulence on the angular coherence of such a system. Additionally, it has been shown in Ref. \cite{Jha:2011dv} that such a measurement could be carried out on either the signal or idler arm in order to obtain the angular coherence properties of the entangled two-photon field produced by spontaneous parametric down conversion \cite{Mair:2001ub,Torres:2003cy}. We expect that angular interference from a double angular-slit holds promise for exploring many interesting quantum phenomena in the high-dimensional space of orbital angular momentum.

This work was supported by the Canada Excellence Research Chairs (CERC) Program, the Natural Sciences and Engineering Research Council of Canada (NSERC), and the DARPA InPho Program.

\bibliographystyle{aps}

\begin{thebibliography}{10}

\bibitem{Mandel:1995}
L.~Mandel, E.~Wolf:
\newblock \emph{{Optical Coherence and Quantum Optics}}:
\newblock Cambridge University Press (1995)

\bibitem{Dixon:2010kz}
P.~B.~Dixon, G.~Howland, M.~Malik, D.~J. Starling, R.~W. Boyd, J.~C. Howell:
\newblock \emph{Physical Review A} \textbf{82}  (2010) 023801

\bibitem{Yao:2006vc}
E.~Yao, S.~Franke-Arnold, J.~Courtial, S.~Barnett, M.~Padgett:
\newblock \emph{Optics Express} \textbf{14}  (2006) 9071

\bibitem{Jack:2008gx}
B.~Jack, M.~Padgett, S.~Franke-Arnold:
\newblock \emph{New Journal Of Physics} \textbf{10}  (2008) 103013

\bibitem{Jha:2008vp}
A.~K. Jha, B.~Jack, E.~Yao, J.~Leach, R.~W. Boyd, G.~S. Buller, S.~M. Barnett,
  S.~Franke-Arnold, M.~J. Padgett:
\newblock \emph{Physical Review A} \textbf{78}  (2008) 043810

\bibitem{Jha:2011dv}
A.~K. Jha, G.~S. Agarwal, R.~W. Boyd:
\newblock \emph{Physical Review A} \textbf{84}  (2011) 063847

\bibitem{Dada:2011dn}
A.~C. Dada, J.~Leach, G.~S. Buller, M.~J. Padgett, E.~Andersson:
\newblock \emph{Nature Physics} \textbf{7}  (2011) 677

\bibitem{Boyd:2011ff}
R.~W. Boyd, A.~Jha, M.~Malik, M.~O'Sullivan, B.~Rodenburg, D.~J. Gauthier:
\newblock \emph{Proceedings of SPIE} \textbf{7948}  (2011) 79480L

\bibitem{Malik:2012ka}
M.~Malik, M.~O'Sullivan, B.~Rodenburg, M.~Mirhosseini, J.~Leach, M.~P.~J.
  Lavery, M.~J. Padgett, R.~W. Boyd:
\newblock \emph{Optics Express} \textbf{20}  (2012) 13195

\bibitem{Wang:2012ha}
J. Wang, J.-Y. Yang, I. M. Fazal, N. Ahmed, Y. Yan, H. Huang, Y. Ren, Y. Yue, S. Dolinar, M. Tur, A. E. Willner: 
\newblock \emph{Nature Photonics} \textbf{6}  (2012) 488

\bibitem{Pors:2011iw}
B.-J. Pors, C. H. Monken, E. R. Eliel, J. P. Woerdman:
\newblock \emph{Optics Express} \textbf{19}  (2011) 6671

\bibitem{DAmbrosio:2012tr}
V. D'Ambrosio, E. Nagali, S. P. Walborn, L. Aolita, S. Slussarenko, L. Marrucci, F. Sciarrino
\newblock \emph{Nature Communications} \textbf{3:961}  (2012)

\bibitem{Mair:2001ub}
A.~Mair, A.~Vaziri, G.~Weihs, A.~Zeilinger:
\newblock \emph{Nature} \textbf{412}  (2001) 313

\bibitem{Torres:2003cy}
J. P.~Torres, A.~Alexandrescu, L.~Torner:
\newblock \emph{Physical Review A} \textbf{68}  (2003) 050301(R)

\bibitem{Zambrini:2006jd}
R.~Zambrini, S.~M. Barnett:
\newblock \emph{Physical Review Letters} \textbf{96}  (2006) 113901

\bibitem{Pires:2010uz}
H.~D.~L. Pires, J.~Woudenberg, M.~P. van Exter:
\newblock \emph{Optics Letters} \textbf{35}  (2010) 889

\bibitem{Groeblacher:2005ec}
S.~Groblacher, T.~Jennewein, A.~Vaziri, G.~Weihs, A.~Zeilinger:
\newblock \emph{New Journal Of Physics} \textbf{8}  (2006) 75

\end{thebibliography}



\end{document}